\documentclass[rnote]{aa} 
\usepackage{graphicx}
\usepackage{natbib}

\usepackage{txfonts}
\begin{document}
 \title{Quiescent X-ray emission from the M9 dwarf LHS 2065}

   \author{J. Robrade
          \inst{1}
          \and
          J.H.M.M. Schmitt\inst{1}
          }


        \institute{Universit\"at Hamburg, Hamburger Sternwarte, Gojenbergsweg 112, D-21029 Hamburg, Germany\\
       \email{jrobrade@hs.uni-hamburg.de}
             }

   \date{Received 07 May 2008; accepted 20 June 2008}

 
  \abstract
{}
   {X-ray emission is an important diagnostics to study magnetic activity in very low mass stars that are presumably fully convective and have an effectively neutral photosphere.}
   {We investigate an archival XMM-Newton observation of LHS 2065, an ultracool dwarf with spectral type M9.}
   {We clearly detect LHS 2065 at soft X-ray energies in less than 1\,h effective exposure time above the 3\,$\sigma$ level with the PN and MOS1 detector. 
No flare signatures are present and we attribute
the X-ray detection to quasi-quiescent activity. From the PN data we derived an X-ray luminosity of $L_{\rm X} = 2.2 \pm 0.7 \times 10^{26}$~erg/s in the 0.3\,--0.8\,keV band,
the corresponding activity level of log~$L_{\rm X}$/$L_{\rm bol}\approx -3.7$ points to a rather active star.
Indications for minor variability and possible accompanying spectral changes
are present, however the short exposure time and poor data quality prevents a more detailed analysis.}
   {LHS 2065 is one of the coolest and least massive stars that emits X-rays at detectable levels in quasi-quiescence, implying the existence of a corona. }
   \keywords{Stars: coronae -- Stars: individual LHS 2065 -- Stars: late-type -- X-rays: stars
               }

   \maketitle
%

\section{Introduction}

The ultracool dwarf star LHS~2065 (GJ~3517) with a spectral type of M9V
is a nearby star located at a distance of 8.6\,pc.
These late-type, very low mass stars are generally assumed to be fully convective and hence a solar-type dynamo is not expected to work.
Further, their cool photospheres should be effectively neutral, leading to a high electric resistivity. 
Therefore the presence of magnetic activity phenomena in the outer atmospheric layers of these stars, especially outside transient events like flares, is remarkable.
X-ray emission, among other diagnostics, can put important constraints on the possible dynamo and activity models.

LHS~2065 is among the latest main-sequence stars detected in X-rays. 
While \cite{fle93} derived an 2\,$\sigma$ upper limit on its X-ray luminosity of
$L_{\rm X}< 3.7 \times 10^{26}$~erg/s from the RASS (ROSAT All Sky Survey) data,
\cite{schmitt02} report a detection of an X-ray flare in 68\,ks pointed observations performed in April/May and Oct./Nov. 1997 with the ROSAT HRI. 
During this event with a decay time of 1.4\,h, a peak luminosity of $L_{\rm X}= 4 \times 10^{27}$~erg/s was derived using a nominal
count to flux conversion factor since the ROSAT HRI had no intrinsic energy resolution. 
In addition to this large flare event, a smaller flare is present in the observations and some indications for
'quiescent' X-ray emission of LHS~2065 were deduced from the absence of clear flaring signal in the light curve.
Specifically, \cite{schmitt02} report non-flaring X-ray emission from LHS~2065
for a period of a few days in autumn 1997 at the level of $L_{\rm X}= 2.6 \times 10^{26}$~erg/s, while for the data taken half a year 
earlier an upper limit of  $L_{\rm X} \le 1.8 \times 10^{26}$~erg/s was derived.
With $L_{\rm bol}=1.2\times 10^{30}$\,erg/s \citep{goli04}, one arrives at an activity level of  log~$L_{\rm X}$/$L_{\rm bol} = -3.7$ in quasi-quiescence, i.e. a level not too far from the 
saturation level around  log~$L_{\rm X}$/$L_{\rm bol} = -3$, while inactive stars like the Sun have log~$L_{\rm X}$/$L_{\rm bol} \approx -7$.  
Consequently, while being X-ray faint in absolute values, LHS~2065 appears to be a rather active star, despite its low temperature of T$_{eff}\lesssim 2400$\,K.
Further evidence for significant magnetic activity on LHS~2065 is its  H${\alpha}$ emission \citep{moh03},
suggesting the existence of a chromosphere, and in its strong magnetic field of B$f >3.9$~kG \citep{rei07}. We note that it was not possible to disentangle the parameters {\it B} and {\it f}
with the applied method that utilizes the profiles of FeH lines.
In view of the rarity of X-ray observations of ultracool dwarfs we examined the
XMM-Newton observations of LHS~2065 and report in this note a clear X-ray detection of LHS~2065 with the EPIC detector
at soft X-ray energies below 1\,keV, despite short effective exposure times and unfavorable high background conditions.
In Sect.\,\ref{ana} we describe the observation and data analysis, in Sect.\,\ref{res} we present our results
and summarize our findings in Sect.\ref{sum}.

\section{Observations and data analysis}
\label{ana}

\begin{figure}[ht]
\includegraphics[width=88mm]{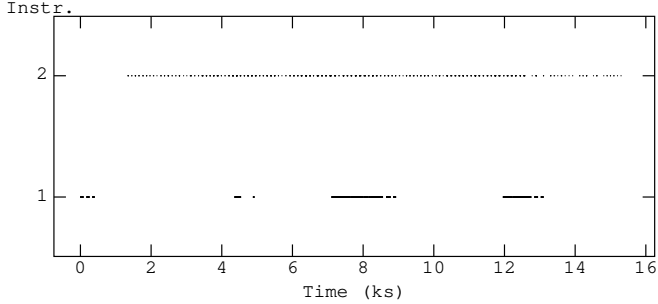}
\caption{\label{obs}Time coverage (event arrival times) during the LHS 2065 observation with the MOS in the {\it lower graph} (Instr.1, on-time: 2.9\,ks) 
and the PN in the {\it upper graph} (Instr.2, on-time: 2.3\,ks) detector.}
\end{figure}

LHS 2065 was observed by {\it XMM-Newton} in April 2004 for approximately 15\,ks (Obs.ID 0203860101); we consider only
data taken with the EPIC (European Photon Imaging Camera) detector, i.e. the MOS and PN, which were both
operated in 'Full Frame' mode with the thin filter. 
Data analysis was carried out with the Science Analysis System (SAS) version~7.1 \citep{sas}.
All detectors are affected by very high background levels, which significantly reduce the effective exposure time 
due to several shutdowns of the MOS detectors and full scientific buffer of the PN detector to less than 3\,ks per instrument. 
Consequently, the time coverage differs for the individual instruments. The MOS data consists of four more or less continuous exposures with long ($\approx$ 20\,--\,40~min) time gaps,
while the PN data consists of one long exposure with short ($\approx$ 1\,min) but very frequent time gaps.
An overview of the temporal coverage is given in Fig.\ref{obs}, where we show the effective exposures of the individual instruments.

To obtain usable datasets for analysis we merged the individual exposures of the MOS detectors,
applied standard selection criteria to the data and constrained our analysis to the respective CCD covering the position of LHS~2065
(the central CCD\,1 in MOS1/2 and CCD\,4 in PN). However, we applied no time filtering, but instead suppressed the background by restricting the energy range of the considered photons since
we expect a possible source signal of LHS~2065 to be predominantly at soft energies below 1\,keV.
The background contamination differs somewhat for the MOS and the PN detector, for the PN we applied overall tighter cuts since its contamination is higher. 
Specifically, we used an energy range of 0.3\,--\,1.0~keV for the MOS and 0.3\,--\,0.8~keV for the PN for our analysis.
We then ran the source detection algorithm 'edetect\_chain' on images created from just these photons, using for the PN a smaller detection cell to stronger focus on the PSF core 
and requiring a maximum likelihood value of at least five. 
All denoted errors are Poissonian errors (1\,$\sigma$ unless otherwise noted) and we consider only detections above the 3\,$\sigma$ level as given by the derived source counts.

\section{Results}
\label{res}

A weak X-ray source is clearly detected at the known position of LHS~2065 in the MOS1 and PN detectors.
The detection at soft X-ray energies, especially in the less contaminated MOS1, is very robust and the 
source is e.g. also detected in the 0.3\,--\,0.8~keV range as used for the PN detector
or in images created only from single pixel (Pattern0) events. The source is not detected at energies above 1.0 keV (e.g. in the 1\,--\,3~keV band) in any detector.
Additionally, also in MOS2 data a photon excess is present in a 20\,\arcsec~circular region centered on LHS~2065 (13\,$\pm$\,8 counts), 
but the source detection algorithm does not accept this excess as a statistically significant source,
possibly due to the triangular shape of its PSF and the smaller excess compared to MOS1 (31\,$\pm$\,9 counts). 
The analyzed observations and source detection parameter are summarized in Table\,\ref{log}. 

The derived source positions from MOS1 and PN agree with each other within errors and are 
within 4\,\arcsec~of the absolute position given by the Simbad database for LHS~2065 for epoch 2004.3 (RA: 08 53 36.052, DEC: -03 29 33.02).
The detected source is the only one present in MOS1, in the PN it is the only source within 5\,\arcmin~from the on-axis position.
From the ROSAT observations we know that no other X-ray source of comparable strength is located in the vicinity of LHS~2065, 
therefore the identification is unambiguous and the soft X-ray spectra makes an unknown extragalactic source very unlikely.

\begin{table} [ht!]
\setlength\tabcolsep{4pt}
\begin{center}
\caption{\label{log} X-ray detection of LHS 2065 with 'edetect\_chain', all data}
\begin{tabular}{l|ll}
\hline
Par. & MOS1 & PN\\\hline
Obs.Time (ks) & 15.6 & 14.0 \\
CCD Livetime (s) & 2869 & 2327 \\
RA  & 08 53 36.397($\pm$ 2.4 \arcsec)  & 08 53 36.096  ($\pm$ 2.8 \arcsec) \\
DEC  & -03 29 36.48 ($\pm$ 2.4 \arcsec) & -03 29 36.47 ($\pm$ 2.8 \arcsec)\\
Counts & 31$\pm$ 9 & 43 $\pm$14\\
Det. likelihood & 10 & 5 \\\hline\hline
\end{tabular}
\end{center}
\end{table}

To investigate whether the excess on the position of LHS~2065 is from a flare, i.e. accumulated during a short period of data-taking, we separated the PN exposure in eight equal time bins
of roughly half an hour each (1750\,s).
However, since the detector is not constantly accumulating data over the full time interval, this results in an average effective exposure time of
291\,s per time bin, with individual bins ranging between 164\,s and 417\,s exposure time.
The source flux is taken from a circular region with a radius of 15\,\arcsec\, around the position of LHS~2065, 
and we account for the PSF wings by requiring preservation of the number of counts as derived by the source detection algorithm. 
We checked two background regions, an adjacent annulus (15\,--\,30\,\,\arcsec) and a larger circular region with a radius of 75\,\arcsec\, slightly offset from the source region,
both leading to very similar results. For the PSF core (FWHM = 6\,\arcsec) of the PN detector we expect in total about five background photons in the exposure, 
for the region used to determine the source flux a mean background level of roughly 50 cts/ks is estimated.
An excess at the position of LHS~2065 is seen in seven out of eight time bins, supporting persistent X-ray emission during the observation,
rather than attributing the signal to a major flare event. 

\begin{figure}[ht]
\includegraphics[width=90mm]{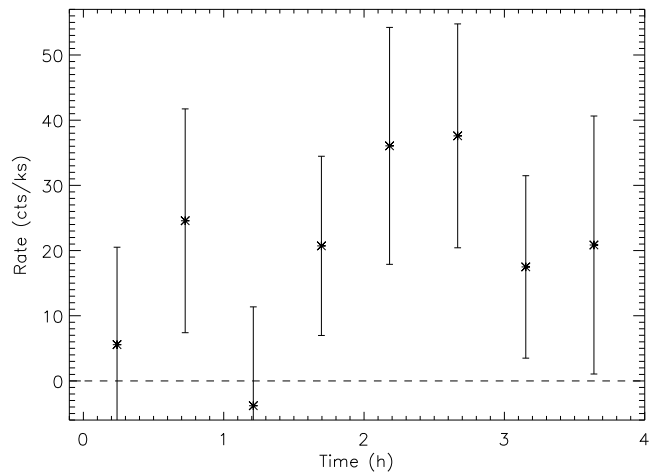}
\caption{\label{pnlc}Light curve of LHS 2065 obtained from PN data in the 0.3\,--\,0.8~keV band with 1\,$\sigma$ errors and using identical absolute time intervals.}
\end{figure}

In Fig.\,\ref{pnlc} we show the thus obtained light curve of LHS~2065, i.e. the background subtracted, exposure time weighted PN count rate in the 0.3\,--\,0.8~keV band for each time bin;
the given errors denote 1\,$\sigma$ statistical errors.
While a constant X-ray flux is consistent with our data given the errors, indications for moderate variability may be present in the derived light curve.
These light curve indicates rather smooth variations, instead of a larger flare as present in the ROSAT data, and it more likely resembles persistent emission. 
Actually, instead of the commonly used term quiescence, the term quasi-quiescence appears to be physically more appropriate,
since variable X-ray flux on all scales (time and amplitude) is generally observed for magnetically active stars.
We finally note, that the count rates derived from MOS1 for the four exposures, i.e. 13.7$\pm$13.9~cts/ks (193\,s),  6.1$\pm$16.2~cts/ks (185\,s) , 
9.9$\pm$4.8~cts/ks (1575\,s), 12.7$\pm$6.4~cts/ks (916\,s), are fully consistent with the PN data.

When converting the observed PN count rate to an energy flux by using ECF's (Energy Conversion Factors) appropriate for the used energy band, we derive a source flux of roughly 
$ 2.5 \pm 0.8 \times 10^{-14}$~erg\,cm$^{-2}$\,s$^{-1}$. This corresponds to an X-ray luminosity of $L_{\rm X}= 2.2 \pm 0.7 \times 10^{26}$~erg/s
and an activity level of log~$L_{\rm X}$/$L_{\rm bol} = -3.7$, close to the value derived from the ROSAT/HRI data outside the obvious flares.
Similar values ($L_{\rm X}= 2.6 \pm 0.8 \times 10^{26}$~erg/s) were obtained for the XMM-Newton PN data, when calculating the source flux with
PIMMS\footnote{http://heasarc.gsfc.nasa.gov/Tools/w3pimms.html} and assuming a thermal plasma around 3\,MK (in the commonly used 0.2\,--\,2.0\,keV band the flux is
roughly 20\% higher).
We note that our errors are by far dominated by the statistical errors, further errors, e.g. in the distance ($8.6\pm 0.1$\,pc) of LHS~2065, were considered in this analysis but do not contribute
significantly.

We finally inspected the energies of the photons extracted from the source region for each time bin.
We find that the source photons show an excess especially at energies between 550\,--\,700\,eV,
which could be attributed the \ion{O}{vii} triplet and \ion{O}{viii} emission around 570\,eV and 650\,eV, 
and that the average photon energy is slightly higher during the X-ray brighter phases (time bins: 2, 5, 6),
pointing to X-ray properties as expected for a coronal signal from a magnetically active star.
A meaningful quantification is difficult due to variability and missing detailed spectral properties of the source and background photons.
The chance probability that the spectral changes are caused by fluctuations in the background photons is estimated to be around 10\% for the most significant bin.
We caution that these findings suffer from low numbers, thus suggesting a longer, deeper 
exposure, which should make it possible to determine the basic quiescent coronal properties of an ultracool dwarf for the first time.

\section{Summary and conclusions}
\label{sum}

   \begin{enumerate}
      \item We detected quasi-quiescent X-ray emission from the ultracool dwarf LHS~2065 with spectral type M9 at soft X-ray energies below 1.0\,keV with a significance
above 3\,$\sigma$ in the PN and MOS1 detector.
The derived X-ray luminosity of $L_{\rm X} = 2.2 \pm 0.7  \times 10^{26}$~erg/s leads
to an activity level of log~$L_{\rm X}$/$L_{\rm bol}\approx -3.7$, pointing to a rather active star.
To our knowledge LHS~2065 is the coolest main sequence star with securely detected quasi-quiescent X-ray emission. 

\item XMM-Newton is well suited to study faint, soft X-ray sources, and
a deeper X-ray observation of LHS~2065 has the potential to put stronger constraints on
the X-ray properties of active stars at the very cool end of the main sequence.

   \end{enumerate}

\begin{acknowledgements}
This work is based on observations obtained with XMM-Newton, an ESA science
mission with instruments and contributions directly funded by ESA Member
States and the USA (NASA). J.R. acknowledges support from DLR under 50QR0803.

\end{acknowledgements}

\bibliographystyle{aa}
\bibliography{/data/hspc44/stch320/Pubs/jansbib}

\end{document}